\documentclass[aps,preprintnumbers,prl,twocolumn,superscriptaddress]{revtex4}

\usepackage{epsfig,latexsym,cancel,amssymb,amsmath}
\usepackage{graphicx}
\usepackage{epstopdf}
\usepackage{hyperref, graphicx, slashed, xcolor, amsmath}
\usepackage{subfigure}
\usepackage{newfile}

\allowdisplaybreaks

\newcommand{\ta}{\text}
\newcommand{\tr}[1]{\text{#1}}

\newcommand{\brr}[1]{\left(#1\right)}
\newcommand{\srr}[1]{\left[#1\right]}

\begin{document}

\title{The BH-PSR Gravitational Molecule} 

\author{Tao Liu}
\email{taoliu@ust.hk}
\affiliation{Department of Physics and Jockey Club Institute for Advanced Study, \\
The Hong Kong University of Science and Technology, Hong Kong S.A.R., China}

\author{Kun-Feng Lyu}
\email{lyu00145@umn.edu}
\affiliation{Department of Physics and Jockey Club Institute for Advanced Study, \\
The Hong Kong University of Science and Technology, Hong Kong S.A.R., China}
\affiliation{School of Physics and Astronomy, University of Minnesota, Minneapolis, MN 55455, U.S.A.}

\begin{abstract}

While axion-clouded black hole (BH) encounters another massive star, a ``gravitational molecule'' (a novel binary system) can form. The axion cloud then evolves at the binary hybrid orbitals, as it occurs at microscopic level to electron cloud in a chemical molecule. To show this picture, we develop a semi-analytical formalism using the method of linear combination of atomic orbitals with the Born-Oppenheimer approximation, and apply it to a BH-PSR (pulsar) gravitational molecule. An oscillating axion-cloud profile and a perturbed binary rotation, together with unique and novel detection signals, are then predicted. Remarkably, the proposed PSR timing and polarization observables, namely the oscillation of periastron time shift and the birefringence with multiple modulations, correlate in pattern, and thus can be properly combined to strengthen the detection.

\end{abstract}

\maketitle

{\bf Introduction} --- ``Superradiance'' of bosons can occur near a spinning black hole (BH)~\cite{Damour:1976kh, Bekenstein:1973mi, Bekenstein:1998nt,Brito:2015oca,Brill:1972xj, Rosa:2009ei, Detweiler:1980uk,Dolan:2007mj,Endlich:2016jgc,Yoshino:2013ofa, Arvanitaki:2010sy}, if their Compton wavelength is comparable to the BH horizon. Then certain energy levels of this BH get populated with these ultralight bosons, by extracting its own angular momentum and energy. Eventually a ``gravitational atom'' with a BH ``nucleus'' forms. Here $|211\rangle$~\cite{Detweiler:1980uk}, a state with gravitational principal, orbital and magnetic quantum numbers $\{n=2,l=1,m=1$\}, may serve as a leading mode.

Among the candidates of ultralight bosons, axion is especially interesting. Axion is originally proposed to solve strong CP problem~\cite{Peccei:1977hh,Peccei:1977ur,Weinberg:1977ma,Wilczek:1977pj,Dine:1982ah,Preskill:1982cy,Abbott:1982af}. This concept is later generalized to string theory~\cite{Svrcek:2006yi,Arvanitaki:2009fg,Cicoli:2012sz}. The axion and axion-like particles (we will not distinguish them below) can serve as a primary component of dark matter in the universe, with a mass ranging from sub eV to $\sim 10^{-22}$~eV~\cite{Marsh:2015xka}.

The advent of multi-messenger astronomy provides a great opportunity to investigate gravitational atoms and explore the physics of ultralight bosons, which has motivated extensive studies in literatures (see, e.g.,~\cite{Irastorza:2018dyq,Brito:2017wnc,Chen:2019fsq,Davoudiasl:2019nlo,Plascencia:2017kca,Isi:2018pzk,East:2018glu,Marsh:2021lqg,Yuan:2021ebu,Baryakhtar:2020gao,Ng:2020ruv,Ng:2020jqd,Kavic:2019cgk,Wen:2021yhz,Chia:2020dye,Siemonsen:2019ebd,Ghosh:2018gaw}). Recently, the study was extended to gravitational-atom (GA) binaries, with their companion being either a BH~\cite{Zhang:2019eid,Fernandez:2019qbj,Baumann:2018vus} or a pulsar (PSR)~\cite{Ding:2020bnl,Kavic:2019cgk}. Especially, the Landau-Zener enegy-level transitions induced by the companion's gravitational perturbation have been suggested to probe for gravitational atoms (or ``gravitational collider'' physics)~\cite{Baumann:2018vus,Baumann:2019eav,Baumann:2019ztm,Ding:2020bnl,Tong:2021whq}.

However, while a GA encounters a BH or a PSR or has such a massive companion, a more generic picture could be the formation of gravitational molecule (GM). Due to orbital hybridization, the axion cloud of GA will flow or partly flow to its companion. This mechanism is different from mass transfer of Type Ia supernova binaries: (1) the axion cloud is generated as a coherent field while the transferred matter in the latter case is particle-like; and (2) the flow can cross the boundary of Roche lobe, where no Newtonian matter transfer takes place. Such a system thus can be viewed as a macroscopic counterpart of a diatomic molecule. 

Although the condensate of the ultralight bosons near a binary has been known~\cite{Ikeda:2020xvt}, analyzing the dynamical evolution of such a GM and its observational phenomenology is highly involved. In this letter we will develop a semi-analytical formalism, inspired by quantum chemistry, to address this problem. We will construct a simplified model for the given GM, analyze its orbital hybridization and state evolution using the method of linear combination of atomic orbitals (LCAO)~\cite{chemlibr} with a Born-Oppenheimer approximation, and then explore its detection signals in astronomy. Such a methodology allows us to study the generic features of GMs without involving much numerical work. We would view it to be a starting point for the more refined analyses later.

For concreteness, we will work on a BH-PSR GM (though the developed analysis formalism can be applied to a BH-BH GM also). The specification of a PSR as the companion is because the BH-PSR binaries are expected to be future precision astronomical laboratories. According to recent estimations~\cite{Chattopadhyay:2020lff,Shao:2018qpt,Shao:2021dbg}, in our galaxy $\sim \mathcal O(10 $ $- 10^3)$ stelar BH-PSR binaries are yet to be discovered. The study of such a GM may benefit from not only its gravitational-wave (GW) signals, but also the PSR timing and polarization signals. Recall, in history the timing signals of PSR B1913+16 have provided the first evidence on GWs~\cite{Taylor:1982zz,Taylor:1989sw,Weisberg:2010zz}. So, a BH-PSR GM will be highly valuable for multi-messenger astronomy.

{\bf Modeling the BH-PSR GM} ---  We model the BH-PSR GM as follows. At $t \leq 0_-$, the axion-clouded BH stays in an initial state of $|211\rangle$, with the PSR's gravitational perturbation being negligibly small. Here the gravitational ``Bohr radius'' and ``fine-structure constant'' of this BH are given by~\cite{Baumann:2019eav}
\begin{eqnarray}
r_c &=& \frac{1}{m_\phi \alpha} = 1.3 \times 10^4 \times \brr{\dfrac{0.015}{\alpha}}  \brr{ \dfrac{10^{-12} \ta{eV}}{m_\phi} }  \ta{km}  \ , \nonumber   \\
\alpha &=& G M_{\rm BH} m_\phi = 0.015 \brr{\dfrac{M_{\rm BH}}{2 M_\odot}} \brr{\dfrac{m_\phi}{10^{-12}\tr{eV}}}   \  .
\end{eqnarray}  
A percent-level $\alpha$ is then obtained for $m_\phi \sim 10^{-12}$~eV and $M_{\rm BH} \sim M_\odot$. In this case, the depletion of the axion cloud by emitting the GWs  occurs at a cosmic time scale. The caused mass loss thus can be neglected. We also assume the BH's angular velocity at its outer horizon to be 
\begin{eqnarray}
\Omega_{\rm BH} <  \frac{E_{211}}{m} \sim m_{\phi} \ .
\end{eqnarray}
Here $E_{211}$ is the $|211\rangle$ eigenenergy. This implies that no more axions will be produced by superradiance. At $t = 0_+$, the axion cloud starts to evolve at the hybrid orbitals of this molecule, where the BH and its companion is separated with $R=R_0$. 

This picture models the diabatic formation of gravitational binaries by, e.g., three-body exchange~\cite{2006tbp..book.....V,Hills175,1975ApJ,1981BAAS13256S,Heggie:2001re,osti_5053454,Ziosi:2014sra,Celoria:2018mzr} (one of the main mechanisms for binary production by which most stellar BH-BH binaries in star clusters may have produced~\cite{osti_5053454,Ziosi:2014sra,Celoria:2018mzr}). Its application can be also extended to the GMs with a prompt orbit shrinking (from $R=R_0'$ to $R=R_0$ with $R_0' > R_0$, at $t=0_+$) caused by either three-body hardening~\cite{2006tbp..book.....V,book-compact_binaries,Celoria:2018mzr} or resonant level transition~\cite{Baumann:2019ztm}, where the axion-could initial state is defined at $R=R_0'$. The binary rotation, which we assume to be circular (with its angular velocity $\omega = |\pmb{\omega}| =\omega_i$ and period $T_b = T_{b,i}$ initially) and anti-parallel to the BH spin here, will be perturbed by the axion-cloud evolution then. Notably, a fully-occupied $|211\rangle$ state has total energy $M_{\rm AC} \sim N m_\phi$ with the occupation number $N \sim [10 f_\phi/( \alpha^{3/2} m_\phi)]^2$ being set by Bosenova limit~\cite{Yoshino:2012kn}. $f_\phi$ is axion decay constant. Assuming $M_{\rm BH} = M_{\rm PSR} =M$ (the BH and the PSR then share the $\alpha$ and $r_c$ values), we have
\begin{eqnarray}
\varepsilon = \frac{M_{\rm AC}}{ M}  \lesssim  \frac{m_e}{m_p} \ll 1   
\end{eqnarray}
for $M \sim M_\odot$ and $f_\phi \lesssim 10^{13}$~GeV at $m_\phi = 10^{-12}$eV~\cite{Arvanitaki:2014wva}. Here $m_e/m_p$ is the mass ratio of electron and proton. This implies that, given the same amount of kinetic energy, the relative change to the BH-PSR motion is much smaller than that of the axion cloud. This analysis thus can be pursued in the spirit of the Born-Oppenheimer approximation in quantum chemistry.  Explicitly, we separate our analysis to two steps. We first analyze the axion-cloud evolution using the LCAO method, assuming the BH-PSR rotation to be  unperturbed. Then we solve the perturbations to the BH-PSR rotation with angular-momentum and energy conservation~\footnote{To converge our discussions in this paper, we will not consider the effects of axion self-gravity. We will also not consider the effects of axion self-interaction except turing on the Bosenova limit where it can play a role~\cite{Yoshino:2012kn}.}.

\begin{figure}[ht]
\centering
\includegraphics[width=6cm]{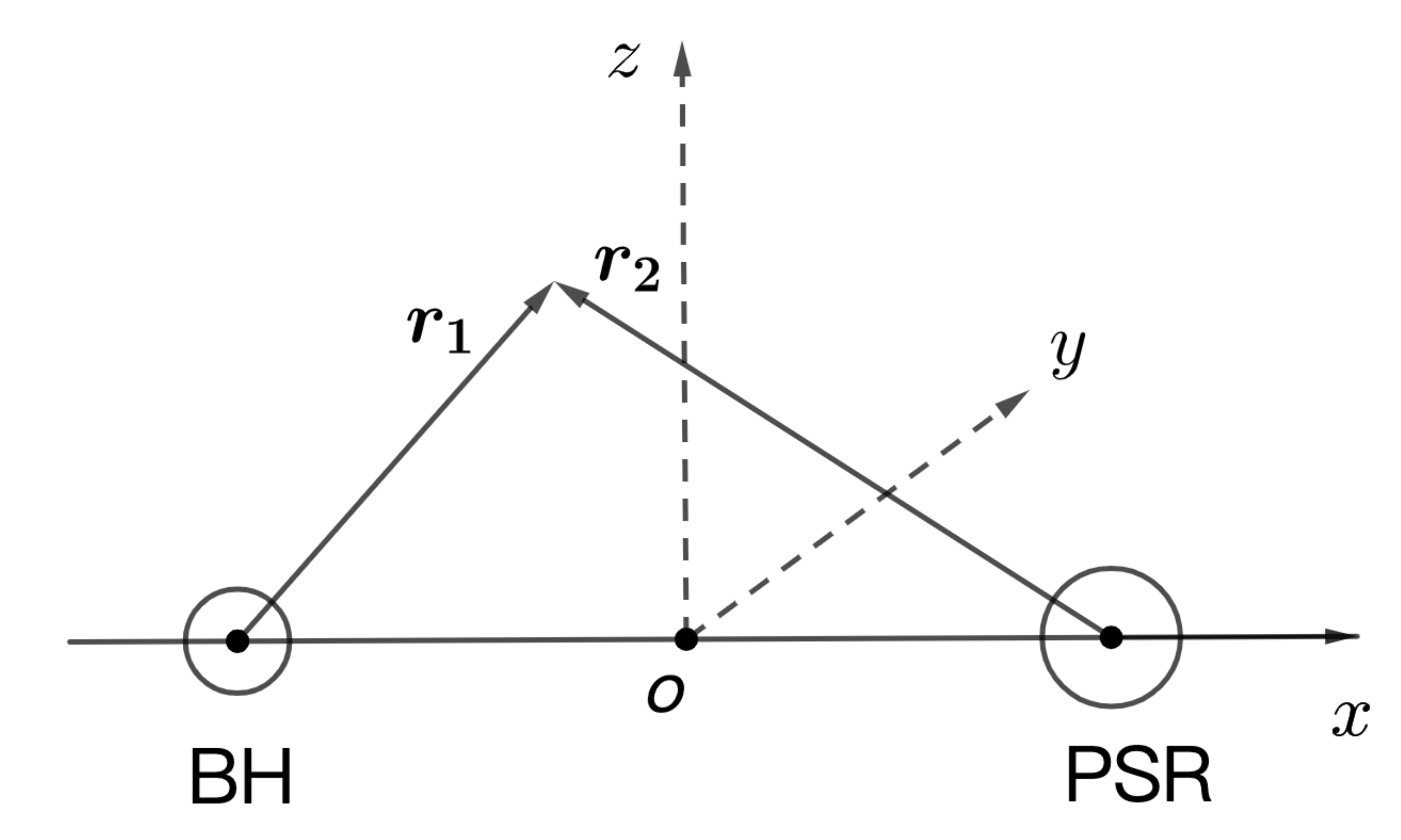}
        \caption{Coordinate system for the rotating reference frame of the BH-PSR GM.}
        \label{fig:coord}
\end{figure}      

The coordinate system for the rotating reference frame of this BH-PSR GM is demonstrated in Fig.~\ref{fig:coord}, where the original point is positioned at the BH-PSR barycenter, and $x$- and $z$-axes are defined by the BH-PSR line and the BH spinning direction, respectively. The axion-cloud profile, as a real scalar field, is described by  
\begin{equation}
\phi(t,{\bf r}) = \dfrac{1}{ \sqrt{2 m_\phi} } \srr{ \psi(t,{\bf r})  \exp{(-i m_\phi t)}  + {\rm h.c.}   } \ .
\end{equation}
Here $\psi$ is complex and satisfies gravitational Schrodinger equation (up to higher-order curvature corrections)
\begin{eqnarray}
i\dfrac{\partial}{\partial t} \psi(t,{\bf r})  = \brr{ \dfrac{ {\bf p}^2}{2 m_\phi}  + V_1 +V_2 + V_c} \psi(t,{\bf r})  \ ,   \label{eq:Sh}
\end{eqnarray}
where $V_{1,2} = - \alpha/|{\bf r_{1,2}}|$ are the BH and the PSR gravitation potentials, and $V_c = - {\pmb \omega} \cdot  {\bf L} = - {\pmb \omega}  \cdot ({\bf r} \times {\bf p})$ is the inertia potential of axion cloud arising from the binary rotation~\cite{2003quant.ph..5081A}. All of them are of order $\sim \mathcal O(\alpha^2)$ in the length unit of $r_c$ (note $\omega =  \brr{2 G M/R^3}^{1/2}=  \sqrt{2} \alpha^2 m_\phi /\brr{R/r_c}^{3/2}$). The Coriolis and centrifugal forces will appear in the Heisenberg equation of ${\bf r}$ then. At last, for $R < 5 r_c$ the radius of Roche lobe (at Lagrange-point-2 side)~\cite{seniorastro-notes} becomes $< 2r_c$, at which the initial axion-cloud distribution is expected to be peaked. Part of the axion cloud might be pushed far away from the binary then. So, our study will focus on the scenario with $R > 5r_c$.

{\bf Orbital Hybridization and State Evolution} --- We introduce a basis for $n_i = 2$ atomic states ($i = 1$, 2 denote ``BH'' and ``PSR'' respectively): 
$|s_{i}  \rangle = | 200_{,i}\rangle$ and     
$|p_{x,i}  \rangle = (| 211_{,i}\rangle   -  | 21-1_{,i} \rangle )/\sqrt{2}$,   
$|p_{y,i} \rangle  = (|211_{,i}\rangle   +  |21-1_{,i} \rangle )/ \sqrt{2} i$.  
Here $|s_{i}  \rangle$ is spherically symmetric in space, while $|p_{x(y),i}  \rangle$ is aligned with $x(y)$-axis. Note, $|p_{z,i}  \rangle$ is irrelevant here since it does not couple with the others via the Hamiltonian. Then we can define a hybrid basis, including the $s_1-s_2$ bond ($\sigma_s$) and anti-bond ($\sigma_s^*$), the $p_x - p_x$ bond ($\sigma$) and anti-bond ($\sigma^*$), and the $p_y - p_y$ bond ($\pi$) and anti-bond ($\pi^*$). Explicitly we have  
 \begin{align}  \label{eq:hb}
 &|\sigma_s \rangle = A_{\sigma_s} \brr{ |s_1 \rangle + | s_2 \rangle  },   \  \  \  |\sigma_s^* \rangle = A_{\sigma_s^*} \brr{ |s_1 \rangle - | s_2 \rangle  }   ,   \\
 &|\sigma \rangle  =  A_{\sigma} \left( | p_{x,1}  \rangle  + |  p_{x,2}  \rangle \right),  \  |\sigma^* \rangle =  A_{\sigma^*} \left( |p_{x,1}  \rangle  -  | p_{x,2}  \rangle  \right), \nonumber \\
 &|\pi \rangle =  A_{\pi} \left(| p_{y,1}  \rangle  +  | p_{y,2}  \rangle \right),  \ |\pi^* \rangle =  A_{\pi^*} \left( |p_{y,1}  \rangle  -  | p_{y,2}  \rangle  \right)  ,   \nonumber 
 \end{align}
with the normalization coefficients  
\begin{equation}
 A_{\sigma_s,\sigma, \pi} = \frac{1}{\sqrt{2 \left( 1 + S_{s,x,y} \right)}}  ,  \  A_{\sigma_s^*,\sigma^*, \pi^*} = \frac{1}{\sqrt{2 \left( 1 - S_{s,x,y} \right)}}  \ .
\end{equation}
Here $S_s = \langle s_{1} |  s_{2}  \rangle$, $S_x = \langle p_{x,1} |  p_{x,2}  \rangle$ and $S_y = \langle p_{y,1} |  p_{y,2}  \rangle$ are overlap integrals of $s$, $p_x$ and $p_y$, respectively. The axion-cloud initial profile is then described by
\begin{eqnarray}
 | 211_{,1}\rangle &=& \frac{\sqrt{2}}{2} (|p_{x,1}  \rangle +i |p_{y,1}  \rangle)  \\ 
 &=&  \frac{1}{2\sqrt{2}}  \left( \frac{|\sigma \rangle}{ A_{\sigma}}   + \frac{|\sigma^* \rangle}{ A_{\sigma}^*} + i \frac{|\pi \rangle}{ A_{\pi}} +i  \frac{|\pi^* \rangle}{ A_{\pi}^*} \right)  \ . \nonumber 
\end{eqnarray} 
Note, both bases are approximately orthogonal only. For the convenience, below we will analyze the orbital hybridization in the hybrid basis and study the observational phenomenology in the atomic basis.

In quantum chemistry, the 2$s$-$2p$ orbital hybridization is often ignored for homonuclear diatomic molecules such as $O_2$ and $F_2$, where the 2$s$-$2p$ energy-level splitting is relatively big. But, this does not apply here, since $ | \langle \sigma_s (\sigma) | V_1 + V_2 | \sigma (\sigma_s) \rangle | >  | \langle \sigma_s | V_1 + V_2 | \sigma_s \rangle  - \langle \sigma | V_1 + V_2 | \sigma \rangle |$. This is reminiscent of some other period-2 diatomic molecules such as $N_2$ and $C_2$, where Coulomb repulsion from the $\sigma$ electrons reduces the 2$s$-$2p$ energy-level splitting and hence enhances its orbital  hybridization. Differently, however, the axion-cloud inertia potential correlates the $\sigma (\sigma^*)$ and $\pi^* (\pi)$ states also, because of $ \langle \pi^* (\pi) | V_c | \sigma (\sigma^*) \rangle  \neq0$. This effect breaks $\sigma$ symmetry (a cylindrical symmetry about $x$-axis) and finally yields a mixing for $\{\sigma_s,\sigma, \pi^* \}$ and $\{\sigma_s^*,\sigma^*, \pi \}$. We denote their respective energy eigenstates as $\xi_{1,2,3}$ and $\xi_{1,2,3}^*$.

\begin{figure}[t]
\centering
\includegraphics[width=6cm]{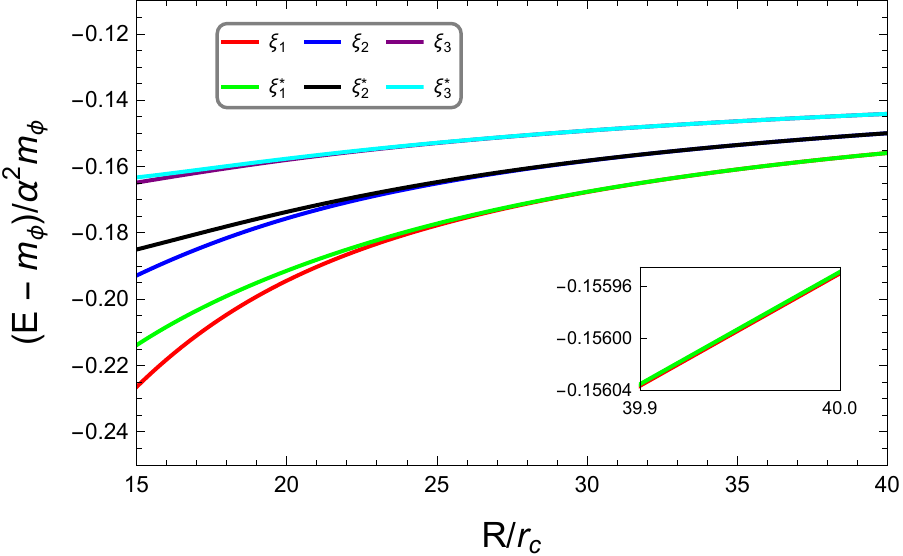}
\caption{Energy eigenvalues of the $n=2$ BH-PSR hybrid states versus the BH-PSR separation.}
\label{fig:eigenE}
\end{figure}

The Hamiltonian matrix for the BH-PSR GM is block-diagonal in the hybrid basis, given by  
$(H_\phi)_{\rm diag}=\alpha^2 m_\phi \times \{ A_\phi, B_\phi \}    + m_\phi \times I$.
Here 
\begin{eqnarray}
     \left .   \begin{array}{lr}
        A_\phi   \\
        B_\phi     
        \end{array}   \right \}  
                      &=& \begin{pmatrix}
-\dfrac{1}{8}-\epsilon & 3\epsilon^2  & 0 \\
3  \epsilon^2 & -\dfrac{1}{8} - \epsilon - 12 \epsilon^3 & i \sqrt{2} \epsilon^{3/2}     \\
0 & - i \sqrt{2}\epsilon^{3/2} & -\dfrac{1}{8}-\epsilon + 6 \epsilon^3
\end{pmatrix}   \nonumber \\ 
&& +\mathcal O( \epsilon^{-3}\exp(-1/2 \epsilon)) 
  \end{eqnarray}
are identical up to an exponentially-suppressed factor.  $\epsilon = \frac{r_c}{R}$ is a perturbation measure for the matrix entries. The eigenvalues are then solved to be 
\begin{equation}
\frac{E_{\xi}  - m_\phi}{\alpha^2 m_\phi} = A \brr {1,1,1} +  \brr {- \sqrt{2} \epsilon^{3\over 2}, 0 ,  \sqrt{2} \epsilon^{3\over 2} } + o(\epsilon^\frac{5}{2})
\end{equation}
with $A= -\frac{1}{8}-\epsilon$. The $\xi_a$ and $\xi_a^*$ profiles and their energy eigenvalues are both $R$-dependent. In the limit of $R\gg r_c$ or $\epsilon \ll 1$, the splitting between $E_{\xi_i^{(*)}}$ and $E_{\xi_j^{(*)}}$ ($i \neq j$) is mainly determined by $\sqrt{2} \epsilon^{3/ 2}$, while the one  between $E_{\xi_i}$ and $E_{\xi_i^{*}}$ is exponentially suppressed. These features are demonstrated in  
Fig.~\ref{fig:eigenE}. The set of $E_{\xi_i^{(*)}}$ define several time scales to characterize this system, including 
\begin{equation}
T_i^\pm = \frac{4 \pi}{|E_{\xi_i} \pm E_{\xi_i^*}|} ,  
\  T_A^\pm = \frac{4\pi}{|(E_{\xi_2} - E_{\xi_3}) \pm (E_{\xi_2^*} - E_{\xi_3^*})|},  
\end{equation}
where $T_A^+ \sim  T_{b,i}$ and $T_i^- $ is exponentially lengthened. 

\begin{figure}[ht]
\centering
\includegraphics[width=8cm]{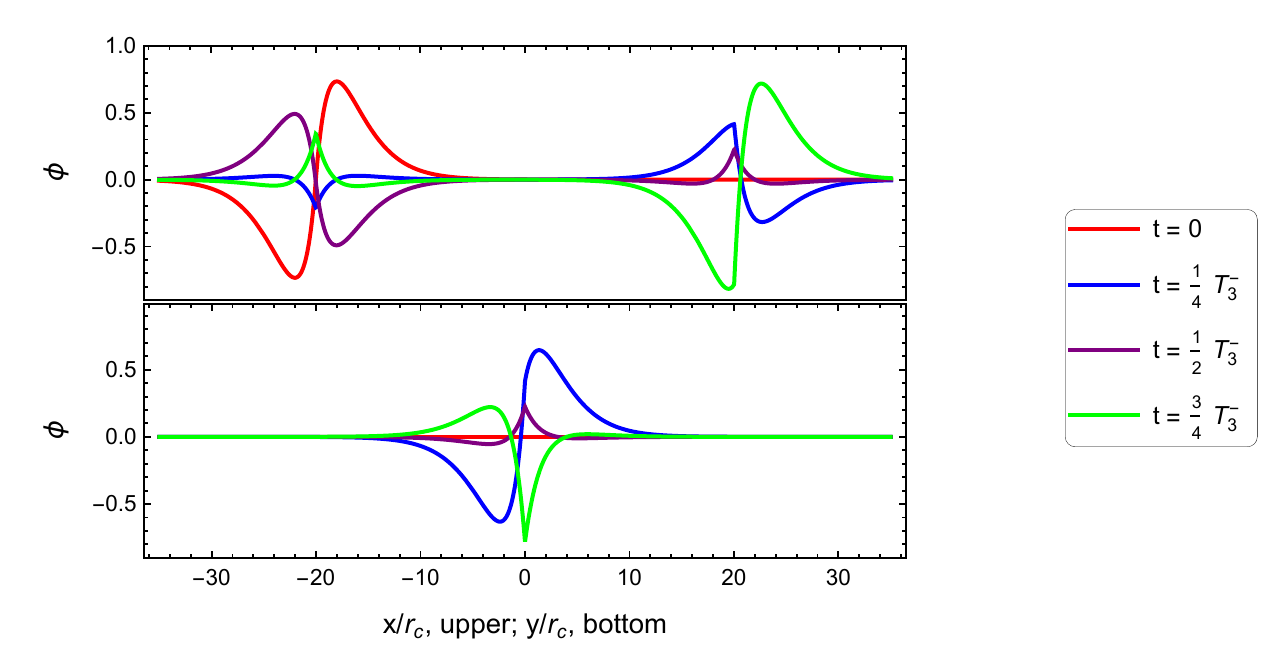}
\caption{Axion-cloud oscillation for the BH-PSR GM, along $x$-axis (upper; $y=z=0$) and $y$-axis (bottom; $x=20 r_c$, $z=0$). }
\label{fig:prof}
\end{figure}

As $| 211_{,1}\rangle$ is not stationary, the axion cloud will evolve in space. In the benchmark scenario defined by Tab.~\ref{tab:BS}, we find (see App.~A for its calculation).
\begin{align}  \label{eq:phir}
 \phi(t,{\bf r}) &= \dfrac{\sqrt{\varepsilon M}}{m_\phi}  \times \\
 & [ 
(  - 0.009 C_1^+ C_1^- - 0.210 C_2^+ C_2^- + 0.218 C_3^+ C_3^- ) s_1  +
\nonumber \\ & 
   (  -0.009 S_1^+ S_1^- - 
  0.210 S_2^+ S_2^-
  + 0.218 S_3^+ S_3^- ) s_2  + \nonumber \\ &   (  0.038 C_1^+ C_1^- 
  - 
  0.002 C_2^+ C_2^- + 
 0.963 C_3^+ C_3^-  ) p_{x,1} + \nonumber \\ &   (  0.038 S_1^+ S_1^-  
 - 
  0.002 S_2^+ S_2^-  + 
 0.963 S_3^+ S_3^-  ) p_{x,2}  + \nonumber \\ &  (  -0.035 S_1^+ C_1^-  
 + 
  0.099 S_2^+ C_2^- + 
0.936  S_3^+ C_3^- ) p_{y,1}  + \nonumber \\ &   
  (  - 0.035 C_1^+ S_1^-
   + 0.100 C_2^+ S_2^- 
   + 0.936 C_3^+ S_3^- ) p_{y,2}  ]  \ ,  \nonumber  
\end{align}
where $S_i^\pm = \sin\brr{\dfrac{2\pi t}{T_i^\pm}}$ and $C_i^\pm = \cos\brr{\dfrac{2\pi t}{T_i^\pm}}$.
At $t = 0_+$, $\phi (t,{\bf r})$ is reduced to $\propto p_{x,1}$. It implies that the axion cloud is distributed near the BH initially. Soon after that, the PSR gets surrounded by the axion cloud massively. As shown in Fig.~\ref{fig:prof}, the axion cloud oscillates between the BH and the PSR with an approximate period of $T_3^-$.

\begin{table}[ht]
\resizebox{0.45\textwidth}{!}{
		\centering
		\begin{tabular}{ |c|c|c|c|c|c|c| }
		\hline
		$M (M_\odot)$  & $ \alpha$ & $m_\phi$(eV) & $f_\phi$(GeV) & $c_\gamma$ & $ r_c$(km) & $R_0(r_c)$    \\	
		\hline 			
	   2  & $0.015$ & $2\times 10^{-12}$  & $5\times10^{12}$&$3 \times 10^3$ & $1.3 \times 10^4$ &  $40$  \\	
	   		\hline
	$h_e$(km) & $T_{b,i}$(s)  & $T_1^-$(s) &$T_2^-$(s) & $T_3^-$(s) & $T_A^+$(s) & $T_A^-$(s) \\	
		\hline 			
	  500 & $3.3\times10^3$ &  $3.1 \times 10^7$ & $3.2 \times 10^7$ & $1.2 \times 10^8$  &  $3.2 \times 10^3$ & $4.4 \times 10^7$  \\		
		\hline
		\end{tabular}
		}
		\caption{Benchmark scenario of the BH-PSR GM.}

		\label{tab:BS}
\end{table}

With the message on the axion-cloud evolution, now we are able to analyze its perturbation to the BH-PSR orbital rotation, by imposing the conditions of angular-momentum and energy conservation. For this purpose, let us define $\beta = (R-R_0)/R_0$ and $\gamma =(\omega -  \omega_i)/\omega_i$ to characterize these effects. During this evolution, the axion-cloud energy is not conserved. First, the transition of axion cloud from the BH $|211\rangle$ state at $t=0_-$ to the binary hybrid states at $t=0_+$ causes a shift to its energy. Second, the axion cloud is not in a stationary state for $t \geq 0_+$ and its energy hence will oscillate as it evolves at the GM hybrid orbitals. Such variations in the axion-cloud energy is interchanged to the binary by gravitational interaction, resulting in non-trivial time dependence for $\beta (t)$ and accordingly for $\gamma (t)$ due to the angular-momentum conservation.    
The axion cloud can interchange its angular momentum with the binary also: 
\begin{equation}
[\hat L_\text{AC}, H_\phi] = \frac{-i \alpha R}{2} \left(\frac{y}{r_1^3} - \frac{y}{r_2^3} \right)  \equiv T_G \neq 0  \    .
\end{equation}
Here $\hat L_\ta{AC} = (x \hat p_y -y \hat p_x)$ is the intrinsic angular-momentum operator of axion cloud along $z$-axis~\footnote{We define ``intrinsic angular momentum'' as the part of angular momentum which is independent of binary rotation. It is different from the angular momentum in the rotating reference frame where the binary rotation can contribute also.}, and $T_G$ is gravitational torque. But, this effect enters at a higher order of $\epsilon$ only. Analytically solving $\beta (t)$ and $\gamma (t)$ in the full time domain is a challenging task. For the purpose of demonstration, let us consider the limit of $t \ll T_A^-$ (note  $T_A^+ \ll T_A^-$) and extend this study to a full time domain later~\cite{LTXZ}. In this limit, 
we have (see App.~B for its derivation) 
\begin{equation}\label{eq:beta}
\begin{split}
\beta (t)  &= -\frac{1}{2} \gamma (t)  \\
&= \pm \varepsilon^{1/2}  \srr{1.1  \times 10^{-2} + 1.9 \cos\brr{\dfrac{2 \pi t}{T_A^+}} }  \ ,
\end{split}
\end{equation}
with $T_A^+  \approx T_b$. Notably, the magnitudes of $\beta$ and $\gamma$ are both $\propto \varepsilon^{1/2}$ at the leading order, which well-justifies the Born-Oppenheimer approximation taken in this study.

{\bf Astronomical Detection} --- The axion-cloud evolution and its perturbation to the binary rotation yield unique and novel signals for detecting the BH-PSR GMs. Here let us consider two PSR observables, namely periastron time shift and birefringence, which are based on its timing and polarization signals respectively. We postpone the study on the GW probe to a later time~\cite{LTXZ}.  

The periastron time shift is a measure of the time dependence of the PSR orbital period, given by~\cite{RevModPhys.66.711,Ding:2020bnl}
\begin{equation}
\Delta t_P(t) = t - T_b(0)\int_{0}^t \dfrac{1}{T_b(t')} dt'  \ .
\end{equation}
It is modulated by the perturbation to the binary rotation, through the $\gamma(t)$ factor in $T_b (t) = T_{b,i}/(1+\gamma (t))$. 

The birefringence~\cite{Carroll:1989vb} occurs when the linearly polarized PSR light travels across an axion field~\cite{Liu:2019brz}, due to the parity-violating interaction: $\mathcal{L} \sim -  \frac{1}{2} g_{\phi\gamma} \phi F_{\mu\nu} \tilde{F}^{\mu\nu}$. Here $F_{\mu\nu}$ is electromagnetic field strength and $\tilde F_{\mu\nu}$ is its dual; $g_{\phi \gamma} = c_\gamma \alpha_{em}/(4 \pi f_a)$ is the Chern-Simons coupling, currently limited to $< \brr{2\times 10^{12} \ta{GeV}}^{-1}$ for $m_\phi \sim 10^{-12}$eV~\cite{Tanabashi:2018oca}; $c_\gamma$ is model-dependent and can vary from $\sim\mathcal O(1)$ to orders of magnitude higher~\cite{Choi:2015fiu,Kaplan:2015fuy,Long:2018nsl,Agrawal:2017cmd}. The PSR polarization position angle (PA) is then varied by~\cite{Harari:1992ea}: 
\begin{equation}
\Delta \Theta = g_{\phi \gamma} \srr{ \phi(t_\ta{r},{\bf r}_\ta{r})  -  \phi(t_\ta{e},{\bf r}_\ta{e}) }   \approx  - g_{\phi \gamma} \phi(t_\ta{e},{\bf r}_\ta{e})   \ .
\end{equation}
Here ``e'' and ``r'' denote the spacetime points of light emission (PSR) and reception (Earth).  
We assume the Earth to be in the $x$-$y$ plane for simplicity. Moreover, for $R \gg r_c$, the contributions of the $s_1$, $p_{x,1}$ and $p_{y,1}$ terms in Eq.~(\ref{eq:phir}) are exponentially suppressed.  
So we have 
\begin{align}  \label{eq:bii}
\Delta \Theta  \approx  &  \brr{\dfrac{c_{\gamma}}{3\times 10^3}}  \brr{\dfrac{M}{2 M_\odot}} \brr{\dfrac{m_\phi}{10^{-12} \tr{eV}}}^2  \brr{\frac{h_e}{500 {\rm km}}}   \nonumber  \\
& \times  \big[ -2.2^\circ S_1^+ S_1^- -  54.8^\circ S_2^+ S_2^-  + 57.1^\circ S_3^+ S_3^-   \nonumber   \\
& + 3.4^\circ  S_3^+ S_3^-   \cos \eta   
- 3.3^\circ  C_3^+ S_3^- \sin \eta \big]  \  ,    
\end{align}
with $\eta = \omega_i \brr{t - \Delta t_P} \sim \frac{2 \pi t}{T_b}$. Here $h_e \sim \mathcal O(10^2) -\mathcal O(10^3)$~km is the emission height of the PSR light~\cite{Mitra:2017sbv}.  This signal is then jointly modulated by the axion-cloud oscillation and the perturbed binary rotation. As a reference, the current  precision of measuring the PSR polarization PA is $\sim \mathcal O(1.0^\circ)$~\cite{Tatischeff:2017qwa,10.1111/j.1365-2966.2009.14935.x,Moran:2013cla}

\begin{figure}[h]
\centering
\includegraphics[width=6cm]{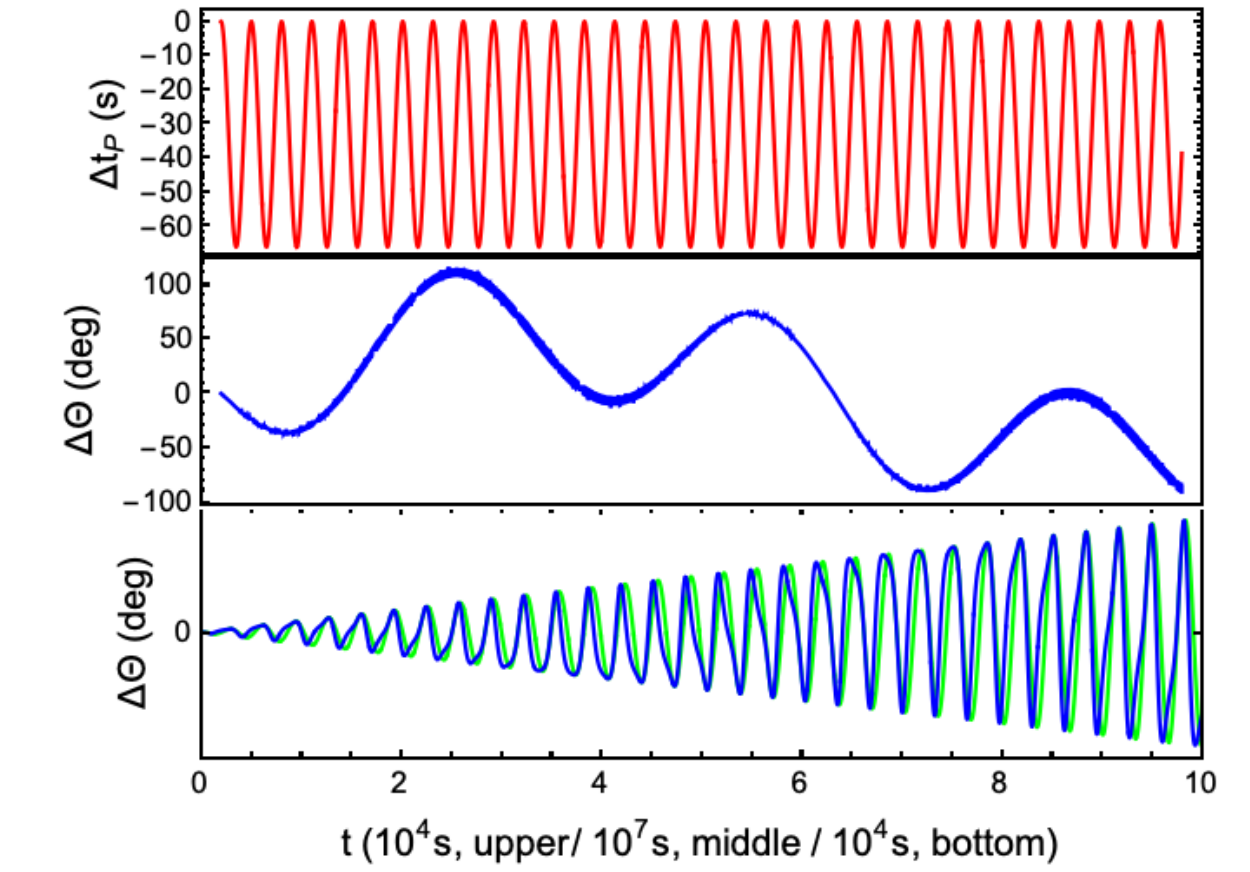}
\caption{Signal patterns of periastron time shift (upper) and birefringence (middle, bottom) for a BH-PSR GM. The bottom panel is a zoom-in of the middle one at $0_+ \leq t \ll T_A^-$, where the $s_2$ contribution is dropped and the $\Delta t_P$ is amplified by 10 times, for a more visual demonstration, and the reference curve (green) is drawn with $\Delta t_P \equiv 0$ ($i.e.$, $\eta = \omega_i t$).}
\label{fig:theta}
\end{figure}

We demonstrate the signal patterns of periastron time shift and birefringence in Fig.~\ref{fig:theta}. $\Delta t_P(t)$ oscillates with a period $T_A^+ \approx T_b$ and an amplitude about tens of seconds. Given the high precision of the PSR timing measurements~\cite{Weisberg:2010zz}, this signal is strong. The $\Delta \Theta$ spectrum is dominated by the $s_2$-mode in Eq.~(\ref{eq:phir}), and hence mainly modulated by $T_2^-$ and $T_3^-$ through the second and third terms in Eq.~(\ref{eq:bii})~\footnote{The $S_2^+$ and $S_3^+$ factors in these two terms oscillate very quickly. Due to the limitation of precision, their observational effects are weak.}. Moreover, a ``hidden'' modulation exists for this spectrum, due to the $\gamma (t)$ dependence of the $\eta$ or $T_b$ parameter. This modulation is manifested as an offset between the blue and green curves in the bottom panel, with a period $\sim T_A^+$ and an amplitude $\sim |\Delta t _P|$. Exactly it is caused by the effect measured with the periastron time shift. These two observables thus can be correlated to strengthen the detection.

{\bf Conclusion and Outlook} --- In this letter, we have developed a LCAO-based formalism to analyze the orbital hybridization of a BH-PSR GM and its state evolution, with the Born-Oppenheimer approximation. An oscillating axion-cloud profile and a perturbed binary rotation are then predicted, together with the unique and novel detection signals. Following this work, we can immediately see several important directions to explore.

First, draw a full picture on the GMs. Despite their rich physics, we consider mainly a BH-PSR GM with $M_{\rm BH}=M_{\rm PSR}$ so far. Generalizing this study to other possibilities, such as a variance of the analyzed one in terms of BH mass ($M_{\rm BH}\neq M_{\rm PSR}$), spin orientation, initial state, orbit eccentricity, curvature corrections, etc., is naturally expected. For example, if the BH and its companion have different mass, the orbital hybridization in more involved, where a dedicated study is necessary.

Second, explore the GM multi-messenger signals. We have proposed two PSR observables, representing its timing and polarization measurements respectively, for detecting the BH-PSR GMs. Complementarily, a GW probe can be applied, where distinguishable signals may arise from the GM evolution. A natural question is then whether a correlation in pattern or some other formats exists between the GW and electromagnetic signals, as it does for the PSR timing and polarization observables. 

At last, we should be aware that the GMs can be a laboratory for studying ultralight bosons other than axions, such as dark photons and massive gravitons. Physics relying on the nature of these particles has been noticed in the studies on gravitational atoms~\cite{Baryakhtar:2017ngi,Brito:2020lup,Caputo:2021efm,Cardoso:2017kgn,Siemonsen:2019ebd}. We leave these interesting topics for future exploration~\cite{LTXZ}. 

\section*{Acknowledgement}

We would thank Ding Pan for discussions on the first-principle calculations, Xuhui Huang, Fu Kit Sheong and Jingxuan Zhang for discussions on the electronic structure of the $H_2^+$ ion (used for our method testing), and Ding Pan, Tixuan Tan, Xi Tong, Yi Wang, Sijun Xu and Kaifeng Zheng for valuable comments on this manuscript. This work is supported by the Collaborative Research Fund under Grant No.~C6017-20G which is issued by the Research Grants Council of Hong Kong S.A.R.

\appendix

\section*{Appendix}

The Appendix contains additional calculation and derivation in support of the results presented in this letter. Concretely, we present in detail the calculation of the axion-cloud profile ($\phi(t,{\bf r})$) in Sec.~A, and the derivation of the axion-cloud perturbation to the binary rotation ($\beta(t)$ and $\gamma(t)$) in Sec.~B.

\section{A. Orbital Hybridization of the BH-PSR GM}

In the hybrid basis, namely $\brr{\sigma_s,\sigma,\pi^*,\sigma_s^*,\sigma^*,\pi}^T$, the Hamiltonian matrix of the considered BH-PSR GM is block-diagonal:
\begin{equation}
\frac{H_\phi - m_\phi \times I}{\alpha^2 m_\phi} =\begin{pmatrix}
A_\phi &   \\
   & B_\phi 
\end{pmatrix}   \  ,
\end{equation}
where  
\begin{equation}
A_\phi = \begin{pmatrix}
-0.150001  &   0.001879 &  0\\
0.001879 & -0.150188 &  -0.005590 i\\
0  &  0.005590 i  & -0.149906
\end{pmatrix} 
\end{equation}
and 
\begin{equation}
B_\phi = \begin{pmatrix}
-0.149999  &   0.001871 &  0\\
0.001871 & -0.150187 &  -0.005590 i\\
0  &  0.005590 i  & -0.149906
\end{pmatrix}  \ .
\end{equation}
The hybrid basis is approximately orthogonal only. By exploiting the Gram-Schmidt orthogonalization process for $A_\phi$ and $B_\phi$, we find their respective eigenvalues and eigenstates to be
\begin{equation}
\begin{split}
\frac{E_\xi - m_\phi}{ \alpha^2 m_\phi } &= (-0.155950,-0.149991,-0.144154)       \\ 
\frac{E_{\xi^*} - m_\phi}{ \alpha^2 m_\phi } &= (-0.155949, -0.149990, -0.144154 )    \ , 
\end{split}
\end{equation}
and
\begin{equation}
V_{\xi} =
\begin{pmatrix}
\xi_1 \\  \xi_2  \\   \xi_3
\end{pmatrix} = \underbrace{
\begin{pmatrix}
-0.22548 &  0.71522  &   -0.66140 i \\
-0.94820  &  -0.00484 & 0.31795 i \\
-0.22422  &  -0.69889  &  -0.67906 i 
\end{pmatrix} }_{U_{A}}
\begin{pmatrix}
\sigma_s \\   \sigma \\  \pi^*
\end{pmatrix}  \label{eq:vxi}
\end{equation}
\begin{equation}
V_{\xi^*} =
\begin{pmatrix}
\xi^*_1 \\  \xi^*_2  \\   \xi^*_3
\end{pmatrix} = 
\underbrace{
\begin{pmatrix}
-0.22529  &  0.71518   &   -0.66163 i \\
-0.94817  &  -0.00474 &  0.31774 i \\
0.22412  &  0.69892  & 0.67930 i 
\end{pmatrix} }_{U_{B}}
\begin{pmatrix}
\sigma_s^* \\   \sigma^*\\    \pi
\end{pmatrix}  \ .        \label{eq:vxis}
\end{equation}

Denoting 
\begin{eqnarray}
V_h = \begin{pmatrix}
\sigma_s \\
\sigma \\
\pi^* 
\end{pmatrix}   \ ,  \quad V_{h^*}  = \begin{pmatrix}
\sigma_s^* \\
\sigma^* \\
\pi
\end{pmatrix}  \  ,
\end{eqnarray}
we can decompose the axion initial state as 
\begin{eqnarray}
 |211,1 \rangle  &=& C^T  V_h +  C^{*T}  V_{h^*}             \nonumber    \\
 &=& C^T U^{-1}_{A}  V_\xi +  C^{*T} U^{-1}_{B} V_{\xi^*}      \ , 
\end{eqnarray}
where
\begin{eqnarray}
C_A^T &=&  \dfrac{1}{2 \sqrt{2}}\begin{pmatrix}
0 & \dfrac{1}{A_\sigma} & \dfrac{i}{A_{\pi^*}} 
\end{pmatrix}
  \ ,   \nonumber \\
C_{A^*}^T  &=& \dfrac{1}{2 \sqrt{2}}\begin{pmatrix}
 0 & \dfrac{1}{A_{\sigma^*}} & \dfrac{i}{A_{\pi}}
\end{pmatrix}
  \ . 
\end{eqnarray}
Then the complex axion field evolves as  
\begin{eqnarray}
\psi(t,{\bf r})  &=&  \sum_{i=1}^3  (C^T U^{-1}_{A})_i  (V_\xi)_i  \exp[-i E_{\xi_i} t]  \\ 
&& +  \sum_{i=1}^3   (C^{*T} U^{-1}_{B})_i (V_{\xi^*})_i  \exp[-i E_{\xi_i^*} t]   \ . \nonumber 
\end{eqnarray}
Replacing $V_{\xi}$ and $V_{\xi^*}$ with Eq.~(\ref{eq:vxi}-\ref{eq:vxis}) and Eq.~(6) in sequence, we finally find Eq.~(12):
\begin{align} 
 \phi(t,{\bf r}) &= \dfrac{1}{ \sqrt{2 m_\phi} } \srr{ \psi(t,{\bf r}) + \psi^*(t,{\bf r})  } = \dfrac{\sqrt{\varepsilon M}}{m_\phi}  \times \\
 & [ 
(  - 0.009 C_1^+ C_1^- - 0.210 C_2^+ C_2^- + 0.218 C_3^+ C_3^- ) s_1  +
\nonumber \\ & 
   (  -0.009 S_1^+ S_1^- - 
  0.210 S_2^+ S_2^-
  + 0.218 S_3^+ S_3^- ) s_2  + \nonumber \\ &   (  0.038 C_1^+ C_1^- 
  - 
  0.002 C_2^+ C_2^- + 
 0.963 C_3^+ C_3^-  ) p_{x,1} + \nonumber \\ &   (  0.038 S_1^+ S_1^-  
 - 
  0.002 S_2^+ S_2^-  + 
 0.963 S_3^+ S_3^-  ) p_{x,2}  + \nonumber \\ &  (  -0.035 S_1^+ C_1^-  
 + 
  0.099 S_2^+ C_2^- + 
0.936  S_3^+ C_3^- ) p_{y,1}  + \nonumber \\ &   
  (  - 0.035 C_1^+ S_1^-
   + 0.100 C_2^+ S_2^- 
   + 0.936 C_3^+ S_3^- ) p_{y,2}  ]  \ .  \nonumber  
\end{align}
Please note that the eigenvalues and eigenstates of $H_\phi$ and hence $\phi(t,{\bf r})$ can be directly solved in the atomic basis of this molecule. Here we work in the hybrid basis first and then convert the solutions to the ones in the atomic basis simply for the convenience of discussions.

\section{B. Perturbation to the BH-PSR Rotation}
\label{app:B}
Below we will analyze the perturbation of the axion-cloud evolution to the BH-PSR rotation, by imposing the conditions of angular-momentum and energy conservation. We will use the Born-Oppenheimer measure, namely $\varepsilon = M_{\rm AC}/M$, as a perturbation parameter, and work in both of the rotating and laboratory reference frames. The Cartesian coordinates for these two frames are denoted as $\{t, x, y, z\}$ and $\{t', x', y', z'\}$, respectively. 
  
At $t=0_+$, the PSR and the BH are located at $\big\{x=\pm \frac{R_0}{2}(1\pm \frac{\varepsilon}{2}), y=z=0 \big\} $, with an angular velocity 
\begin{equation}
\omega_i = \omega (t =0_+) = \dfrac{1}{R_0} \brr{\dfrac{2 G M}{R_0}}^{1/2} \brr{1 + \dfrac{\varepsilon}{4}  }  \ .
\end{equation}
The BH-PSR binary thus has an initial orbital angular momentum 
\begin{eqnarray} 
L_{b,i} &=& M \bigg\{ \srr{\dfrac{R_0}{2}\brr{1+\dfrac{\varepsilon}{2}}}^2  + \srr{\dfrac{R_0}{2}\brr{1-\dfrac{\varepsilon}{2}}}^2 \bigg\} \omega_i  \nonumber \\
 &=& M \brr{\dfrac{G M R_0}{2}}^{1/2}  \brr{1 + \dfrac{\varepsilon}{4}  } \\
 &\propto&  \frac{M}{m_\phi}  \brr{1 + \dfrac{\varepsilon}{4}  }  \nonumber  \ .
\end{eqnarray}
The first term here represents a leading-order contribution to the total angular momentum. 

The axion cloud has two contributions to the total angular momentum. Both of them are of the order of $\mathcal O(\varepsilon)$. The first one is from the axion-cloud rotation around the binary mass center, and hence intrinsic. It is given by  
\begin{equation}
L_\ta{AC} = - \int d^3 \vec{x} \dfrac{\partial \phi}{\partial t} \brr{x \dfrac{\partial}{\partial y} - y \dfrac{\partial}{\partial x}} \phi =  \dfrac{\varepsilon M}{m_\phi}  X_\ta{AC}(t)  \ ,    \label{eq:LAC}
\end{equation}
where 
\begin{eqnarray}
X_\ta{AC}  &\approx&
\dfrac{1}{m_\phi} \bigg[  0.451 \brr{E_{\xi_3}  + E_{\xi_3^*} } 
+ 0.024 \brr{E_{\xi_2} + E_{\xi_3}}  \nonumber \\&& \times \cos\brr{\dfrac{4 \pi t}{T_{23}^-}} 
 + 0.024 \brr{E_{\xi_2^*} + E_{\xi_3^*}} \cos\brr{\dfrac{4 \pi t}{T_{23}^{*-}}} 
 \bigg]    \nonumber \\
&=  & 0.901  
+ 0.047 \cos\brr{\dfrac{4 \pi t}{T_{23}^-}} 
 + 0.047 \cos\brr{\dfrac{4 \pi t}{T_{23}^{*-}}}  \nonumber \\
&=    & 0.901  
+ 0.094 \cos\brr{\dfrac{4 \pi t}{T_A^+}}   \cos\brr{\dfrac{4 \pi t}{T_A^-}}   \  ,  
 \label{eq:Xacp}
\end{eqnarray}
with $T_{ij}^\pm = 4\pi/|E_{\xi_i} \pm E_{\xi_j}|$ and $T_{ij}^{*\pm} = 4\pi/|E_{\xi_i^*} \pm E_{\xi_j^*}|$. In deriving Eq.~(\ref{eq:Xacp}) we have left out the terms which are either suppressed by the coefficients of the $T_1^\pm$-related terms in Eq.~(12) or are proportional to the energy gap $|E_{\xi_i^{(*)}}  - E_{\xi_j^{(*)}}|$. We have also applied $E_{\xi_i^{(*)}} \approx m_\phi$ to obtain the result in the last line.

The second contribution of the axion cloud is induced by the binary rotation. It is given by
\begin{equation}
L_\ta{AO} =\omega  \int d^3 \vec{x}  \srr{\brr{x \dfrac{\partial}{\partial y} - y \dfrac{\partial}{\partial x}} \phi}^2 
=  \dfrac{\varepsilon M}{m_\phi}  X_\ta{AO}(t)
  \ ,    \label{eq:LAO}
\end{equation}
where
\begin{eqnarray}
	 X_\ta{AO}(t) &=&  10^{-7} \times \bigg[ 499.9 \nonumber\\
	&& +32.5  \cos\brr{\dfrac{4 \pi t}{T_{A}^-}} \cos\brr{\dfrac{4 \pi t}{T_{A}^+}}  \nonumber\\
	&& + 33.6 \cos\brr{\frac{(E_{\xi_1}-E_{\xi_1^*}+E_{\xi_3}-E_{\xi_3^*})t}{2}}  \nonumber\\
	&& \times \cos\brr{\frac{(E_{\xi_1}+E_{\xi_1^*}+E_{\xi_3}+E_{\xi_3^*})t}{2}}  \nonumber\\
	&& - 192.9 \cos\brr{\dfrac{4\pi t}{T_3^-}}\cos\brr{\dfrac{4\pi t}{T_3^+}}  \nonumber\\
	&& - 109.9 \cos\brr{\frac{(E_{\xi_2}-E_{\xi_2^*}+E_{\xi_3}-E_{\xi_3^*})t}{2}} \nonumber\\
	&& \times \cos\brr{\frac{(E_{\xi_2}+E_{\xi_2^*}+E_{\xi_3}+E_{\xi_3^*})t}{2}}
	\bigg]   \ .  \label{eq:Xaop}
\end{eqnarray}

As a consistency check, let us combine $L_\ta{AC}$ in Eq.~(\ref{eq:LAC}) and $L_\ta{AO}$ in Eq.~(\ref{eq:LAO}). This yields exactly the total angular momentum of the GM in the laboratory reference frame
\begin{eqnarray}
 L_\ta{AC} + L_\ta{AO} &=&  \int d^3 \vec{x}  \brr{\dfrac{\partial}{\partial t} - \omega x \dfrac{\partial}{\partial y} + \omega y \dfrac{\partial}{\partial x}}   \phi   \nonumber \\ && \times  \brr{ - x \dfrac{\partial}{\partial y} + y \dfrac{\partial}{\partial x}} \phi    \nonumber \\
 &=&   \int d^3 \vec{x'} \dfrac{\partial \phi}{\partial t'} \brr{- x' \dfrac{\partial}{\partial y'} + y' \dfrac{\partial}{\partial x'}} \phi  \ .
\end{eqnarray} 
Compared to $L_\ta{AC}$, however, $L_\ta{AO}$ suffers double suppressions caused by non-relativistic rotation of the binary and non-relativistic nature of $\phi$ ($i.e.$, $\Big |\dfrac{\partial \phi}{\partial x} \Big|$, $\Big|\dfrac{\partial \phi}{\partial y}\Big| \ll \Big|\dfrac{\partial \phi}{\partial t}\Big|$). So its magnitude is many orders of magnitude smaller than that of $L_\ta{AC}$.

The axion-cloud energy can be decomposed into intrinsic and induced parts also. It is then given by 
\begin{eqnarray}
E_\text{AC}(t) &=& \int d^3 \vec{x'} \Bigg [\dfrac{1}{2} \brr{\dfrac{\partial \phi}{\partial t'}}^2 + \dfrac{1}{2} \brr{\nabla' \phi}^2   \nonumber \\ 
&&  + \dfrac{1}{2} m_\phi^2 \phi^2 + V(\phi) \Bigg ] \nonumber \\
&=& \dfrac{\varepsilon M}{m_\phi}  \langle E_\phi \rangle (t) - \int d^3 \vec{x} \,
\omega \, \dfrac{\partial \phi}{\partial t} \brr{ -x \dfrac{\partial \phi}{\partial y} + y \dfrac{\partial \phi}{\partial x} } \nonumber \\
&&+  \int d^3 \vec{x} \dfrac{\omega^2}{2}  \brr{ -x \dfrac{\partial \phi}{\partial y} + y \dfrac{\partial \phi}{\partial x} }^2  \nonumber \\
&=&  \dfrac{\varepsilon M}{m_\phi} \bigg[ \langle E_\phi \rangle (t)   -  \omega X_\ta{AC}(t) + \dfrac{\omega}{2} X_\ta{AO}(t) \bigg] \ .
\end{eqnarray}
Here 
\begin{eqnarray}\label{eq:Ephi_t}
 \langle E_\phi \rangle (t)   &=&  \dfrac{m_\phi} {\varepsilon M} \int d^3 \vec{x} 
    \Bigg [\dfrac{1}{2} \brr{\dfrac{\partial \phi}{\partial t}}^2 + \dfrac{1}{2} \brr{\nabla \phi}^2 
    \nonumber \\ && + \dfrac{1}{2} m_\phi^2 \phi^2 + V(\phi) \Bigg ] \nonumber \\
    &=&m_\phi \bigg[ 1 - 6.6\times 10^{-5} \nonumber \\
    && - 5.9 \times 10^{-8} \cos\brr{E_{\xi_2}  - E_{\xi_3} }t \nonumber\\
    && - 5.9 \times 10^{-8}  \cos\brr{E_{\xi_2^*}  - E_{\xi_3^*}} t   \bigg]
\end{eqnarray}
is intrinsic, with $V(\phi) = -G M m_\phi^2 \phi^2 \brr{r_1^{-1} + r_2^{-1}}$, while $\omega X_\ta{AC}(t)$ and $\omega X_\ta{AO}(t)$ are induced by binary rotation. The factor $\dfrac{\varepsilon M}{m_\phi} = \dfrac{M_{\rm AC}}{m_\phi}$ denotes the occupation number of axion in its initial state. In this calculation, the cosine factors oscillating with a short period, namely a period $\sim m_\phi^{-1}\sim \mathcal O(10^{-3})$s, have been dropped.

With these inputs, we are able to write down the condition of angular-momentum conservation
\begin{eqnarray}
&& L_{b,i} +   \dfrac{\varepsilon M}{m_\phi} (X_{{\rm AO},i} - X_{{\rm AC},i})  \nonumber  \\
&=& \frac{1}{2} M R^2\omega  +   \dfrac{\varepsilon M}{m_\phi} (X_{{\rm AO}}  - X_{\rm AC})  \ , \label{eq:amc}
\end{eqnarray}
and the condition of energy conservation
\begin{eqnarray}
&& \dfrac{1}{4} M R_0^2 \omega_i^2 - \dfrac{GM^2}{R_0} +  \nonumber   \\
&&\varepsilon \dfrac{M}{m_\phi} \bigg[ \langle E_\phi \rangle_i   - 
 \omega_i X_\ta{AC}(t_i) + \dfrac{\omega_i}{2} X_\ta{AO}(t_i) \bigg]
\nonumber \\
&=&\frac{1}{4} M (\dot R^2 + R^2 \omega^2) - \dfrac{G M^2}{R} +  \nonumber  \\
&&
\varepsilon \dfrac{M}{m_\phi} \bigg[ \langle E_\phi \rangle (t)   -  \omega X_\ta{AC}(t) 
 + \dfrac{\omega}{2}  X_\ta{AO}(t) \bigg] 
 \label{eq:ec}
\end{eqnarray}
at the order of $\mathcal O(\varepsilon)$, for the GM evolution. The BH spin couples to these two equations at a high order of $\alpha$ only and hence has been neglected here.
Notably, the transition of axion cloud from the BH (or GA) $|211\rangle$ state at $t=0_-$ to the binary (or GM) hybrid states at $t=0_+$ causes a shift to its energy (recall $E_{211}  = (1 - \alpha^2/4) m_\phi$)
\begin{equation}
\dfrac{\varepsilon M}{m_\phi} \brr{ \langle E_\phi \rangle (t=0_+) -  E_{211}} = \varepsilon M \brr{\frac{\alpha^2}{8}   - 6.6 \times 10^{-5}} < 0 \ . 
 \end{equation} 
The reduced energy will be transferred to the binary gravitationally and should be included in this energy conservation condition. So we define $\langle E_\phi \rangle_i$ here to be  
\begin{eqnarray} 
\langle E_\phi \rangle_i  = E_{211}  \ .  \label{eq:Ephii}
\end{eqnarray}

These two equations can be rewritten as
\begin{equation}\label{eq:dottheta}
\omega =  \brr { \frac {M R^2}{2}}^{-1} \left[ \frac{M R_0^2}{2} \omega_i    + \frac{\varepsilon M }{m_\phi}      ( \Delta X_{\rm AC} -  \Delta X_{\rm AO}) \right ] 
\end{equation}
and 
\begin{equation}
\dot{\beta}^2 + \dfrac{2 G M}{R_0^3} \beta^2 
= \dfrac{2 \varepsilon}{m_\phi R_0^2} \srr{ \brr{\dfrac{2 G M}{R_0} }^{1 / 2} \Delta X_{\rm AO}  -2 \Delta \langle E_\phi \rangle (t) }   \ ,
\end{equation}
where $\Delta X_{\rm AC}= X_{\rm AC} (t) - X_{{\rm AC},i}$, $\Delta X_{\rm AO}= X_{\rm AO} (t) - X_{{\rm AO},i}$, $\Delta \langle E_\phi \rangle (t) = \langle E_\phi \rangle (t) -\langle E_\phi \rangle_i$ and $\beta(t) = \frac{R}{R_0} - 1$. With $\langle E_\phi \rangle (t)$ calculated in Eq.~(\ref{eq:Ephi_t}) and $\langle E_\phi \rangle_i$ defined in Eq.~(\ref{eq:Ephii}), $\Delta \langle E_\phi \rangle (t)$ is found to be  
\begin{eqnarray}
\Delta \langle E_\phi \rangle (t) &\approx & m_\phi \Bigg[ -9.8 \times 10^{-6}   \nonumber \\
&& - 1.2 \times 10^{-7} \cos \brr{\dfrac{2 \pi t}{T_A^+} } \cos \brr{\dfrac{2 \pi t}{T_A^-} } \Bigg] \ .
\end{eqnarray}	
Finally, in the limit of $t \ll T_A^-$ (note  $T_A^+ \ll T_A^-$), we have
\begin{equation}
\beta (t) =  \pm \varepsilon^{1/2}  \srr{1.1  \times 10^{-2} + 1.9 \cos\brr{\dfrac{2 \pi t}{T_A^+}} }  \ ,  \label{eq:beta}
\end{equation}
and
\begin{equation}
\gamma(t) = \dfrac{\omega(t)}{\omega_i} - 1 = -2 \beta(t) + \mathcal O(\varepsilon)  \ .
\end{equation}
Here the relation 
\begin{eqnarray}
T_A^+ =2 \pi \bigg ( \frac{2 m_\phi R_0^3}{\alpha}  \bigg)^{1/2} 
\end{eqnarray}
has been applied.

\bibliography{references}


\end{document}